\documentclass[
 amsmath,amssymb,
 aps,
]{revtex4-1}

\usepackage{graphicx}
\usepackage{dcolumn}
\usepackage{bm}

\begin{document}

\title{ BF theory explanation of the entropy for nonrotating isolated horizons}
\author{Jingbo Wang}
\email{ shuijing@mail.bnu.edu.cn}
\affiliation{ Department of Physics, Beijing Normal University,\\ Beijing, 100875, China}
\author{Yongge Ma}
\email{Correspondingauthor:mayg@bnu.edu.cn}
\affiliation{ Department of Physics, Beijing Normal University,\\ Beijing, 100875, China}
\author{Xu-An Zhao}
\email{zhaoxa@bnu.edu.cn}
\affiliation{ School of Mathematical Sciences, Beijing Normal University,\\ Beijing, 100875, China}
 \date{\today}
\begin{abstract}
We consider the nonrotating isolated horizon as an inner boundary of a four-dimensional asymptotically flat spacetime region. Due to the symmetry of the isolated horizon, it turns out that the boundary degrees of freedom can be described by a SO(1,1) BF theory with sources. This provides a new alternative approach to the usual one using Chern-Simons theory to study the black hole entropy. To count the microscopical degrees of freedom with the boundary BF theory, the entropy of the isolated horizon can also be calculated in the framework of loop quantum gravity. The leading-order contribution to the entropy coincides with the Bekenstein-Hawking area law only for a particular choice of the Barbero-Immirzi parameter, which is different from its value in the usual approach using Chern-Simons theory. Moreover, the quantum correction to the entropy formula is a constant term rather than a logarithmic term.
\end{abstract}
\maketitle

\section{Introduction}
While the notion of the event horizon of a black hole is based on the global structure of the spacetime\cite{he1}, the notion of an isolated horizon is defined quasilocally as a portion of the event horizon which is in equilibrium\cite{ash3}. As expected, the laws of black hole mechanics can be generalized to those of an isolated horizon\cite{ash2,ash3}. The advantage of the quasilocal notion of an isolated horizon is that it allows us to explore the statistical mechanical origin of its entropy by some local
quantum gravity theory. In fact, various attempts have been made in the framework of loop quantum gravity (LQG)\cite{rov1,thie1,ash0,ma1} to account for the entropy of the isolated horizon\cite{asha,ash1}. In the usual treatment the degrees of freedom of the isolated horizon are described by Chern-Simons theory with the SU(2)\cite{kaul3,kaul4,kaul5,enpp1,enpp2} [or U(1)\cite{ash1,asha,ash4}] gauge group. The relation between the approaches with the two different gauge groups was discussed in Refs.\cite{kaul1,kaul2}. For a recent review on the entropy of the isolated horizon in LQG, we refer to Refs.\cite{blv,el1,dpp1}.

Although it is feasible to account for the entropy of an isolated horizon using the boundary Chern-Simons theory, this approach cannot be valid for arbitrary dimensions of the horizon since Chern-Simons theories are only well defined on odd-dimensional manifolds. The aim of this paper is to use BF theory--another topological field theory--to account for the entropy of the isolated horizon in the framework of LQG, which admits the possibility of applying the theory to an arbitrary-dimensional horizon. Note that a tentative attempt to describe the horizon boundary degrees of freedom using BF theory was first made in Ref.\cite{hus1}. As the first step, we consider the nonrotating isolated horizon in four-dimensional spacetime. We will show that, with the boundary condition for the isolated horizon, the horizon degrees of freedom can be described by a SO(1,1) BF theory, which is well defined on an arbitrary-dimensional manifold. Hence, in this case, the entropy of the isolated horizon can also be counted using the boundary BF theory in the framework of LQG.

This alternative approach gives an entropy formula that is different from that given by Chern-Simons theory. Thus the two approaches indicate different values of the Barbero-Immirzi parameter.

The paper is organized as follows. In Sec.II, the covariant phase-space method, we derive the symplectic structure for the spacetime with a nonrotating isolated horizon as an inner boundary. The presymplectic form can be split into the bulk term and the boundary term. In Sec.III, we identify the boundary degrees of freedom with those of the BF theory. We quantize BF theory with sources and give the corresponding Hilbert space. In Sec.IV, we set up the boundary condition to relate boundary fields to the bulk fields and calculate the entropy of the isolated horizon. In light of LQG, the Bekenstein-Hawking area law of black hole entropy is obtained. Our results are discussed in Sec.V.

\section{The Symplectic Structure}
Let us first consider a four-dimensional spacetime region $\mathcal{M}$ with an isolated horizon $\Delta$ as an inner boundary. As in the usual treatment in LQG\cite{ash1}, we are going to employ the covariant phase-space method\cite{lw1,cps1} to derive the symplectic structure of the system.

The Palatini action of general relativity on $\mathcal{M}$ reads
\begin{equation}\label{1}\begin{split}
    S[e,A]=-\frac{1}{4\kappa}\int_{\mathcal{M}} \varepsilon_{IJKL} e^I\wedge e^J \wedge F(A)^{KL}\\+\frac{1}{4\kappa}\int_{\tau_{\infty}} \varepsilon_{IJKL} e^I\wedge e^J \wedge A^{KL},
\end{split}\end{equation}
where we let $\kappa\equiv 8\pi G$, $e^I$ is the co-tetrad, $A^{IJ}$ is the SO(3,1) connection 1-form, and $F(A)^{KL}\equiv dA^{KL}+[A,A]^{KL}$ is the curvature of the connection $A^{KL}$. For convenience, we define the solder form $\Sigma^{IJ}\equiv e^I\wedge e^J$ and its dual $(*\Sigma)_{KL}=1/2\varepsilon_{IJKL} \Sigma^{IJ}$. All fields will be assumed to be smooth and satisfy the standard asymptotic boundary condition at infinity, $\tau_{\infty}$. The boundary term at $\tau_{\infty}$ is required for the differentiability of the action.
From the first variation of the action (\ref{1}) we can get the symplectic potential density,
\begin{equation}\label{2}
    \theta(\delta)=\frac{1}{2\kappa}(*\Sigma)_{IJ}\wedge \delta A^{IJ}.
\end{equation}
Thus the second-order variation will give the presymplectic current,
\begin{equation}\label{2a}
    J(\delta_1,\delta_2)=\frac{1}{\kappa}\delta_{[1}(*\Sigma)_{IJ}\wedge \delta_{2]} A^{IJ}.
\end{equation}
The variational principle implies $d J=0$. Applying Stokes' theorem to the integration $\int_{\mathcal{M}} dJ=0$, we can get the following equation:
\begin{equation}\label{2b}\begin{split}
   \frac{1}{\kappa} (\int_{M_1}\delta_{[1}(*\Sigma)_{IJ}\wedge \delta_{2]} A^{IJ}-\int_{M_2}\delta_{[1}(*\Sigma)_{IJ}\wedge \delta_{2]} A^{IJ}\\+\int_{\Delta}\delta_{[1}(*\Sigma)_{IJ}\wedge \delta_{2]} A^{IJ})=0,
\end{split}\end{equation}
where $M_1,M_2$ are space-like boundaries of $\mathcal{M}$. Note that the boundary integral at spatial infinity $\tau_{\infty}$ vanishes by suitable fall-off conditions \cite{el1}. Next we will show that the horizon integral in Eq.(\ref{2b}) is a pure boundary contribution, i.e, the symplectic flux across the horizon can be expressed as a sum of two terms corresponding to the two-sphere $H_1=\Delta \cap M_1$ and $H_2=\Delta \cap M_2$.

To describe the geometry near the isolated horizon, it is convenient to employ the Newman-Penrose formalism with the null tetrad $(l, n, m, \bar{m})$\cite{np1}. Let the real vectors $l$ and $n$ coincide with the outgoing and ingoing future-directed null vectors at the horizon $\Delta$, respectively. For the nonrotating isolated horizon which we are considering, the components $\pi$ and $\bar{\pi}$ of $l^a\nabla_a n$ along the complex null vectors $\bar{m}$ and $m$, respectively are vanishing on $\Delta$\cite{wxn}. In the neighborhood of $\Delta$,  we choose the Bondi coordinates given by $(v,r,x^i),\,i=1,2$, where the horizon is given by $r=0$\cite{wxn,ksh}. The fields can be expanded in a power series in the coordinate $r$ away from the horizon. Acting on the function, the null tetrad in the neighborhood can be written as
 \begin{equation} \label{2c}
\left\{ \begin{aligned}
         n^a \nabla_a &= -\frac{\partial}{\partial r} \\
         l^a \nabla_a &= \frac{\partial}{\partial v}+U\frac{\partial}{\partial r}+X^i \frac{\partial}{\partial x^i}\\
         m^a \nabla_a &=\Omega\frac{\partial}{\partial r}+\xi^i\frac{\partial}{\partial x^i}
        \end{aligned} \right.
\end{equation}
where the frame functions U and $X^i$ are real, while $\Omega$ and $\xi^i$ are complex functions of $(v,r,x^i)$.

Near the horizon, up to the second order of $r$, the metric components can be written as\cite{wxn,ksh}
\begin{equation}\label{39}\begin{split}
 g^{rr}=2 (\tilde{\kappa} r+Re(\Psi_2^{(0)}) r^2),\,g^{vr}=1,\\
g^{ri}=4Re(1/2\Psi_3^{(0)}\xi^i_{(0)})r^2,\,g^{ij}=\xi^i \bar{\xi}^j+\bar{\xi}^i \xi^j,
\end{split}\end{equation}
where $\tilde{\kappa}$ is the surface gravity on the horizon, $\Psi_i$ are the components of the Weyl tensor, and the subscript $(0)$ denotes taking values on $\Delta$.  In the following, if it is not specified, the functions in front of $r$ are all functions of $(v,x^i)$. In the coordinate neighborhood the null co-tetrad can also be written up to the second order of $r$ as\cite{ksh}

\begin{equation} \label{40}
\left\{ \begin{aligned}
         n = -dv, \\
         l=dr-(\tilde{\kappa}r+Re(\Psi_2^{(0)})r^2)dv-Re(\Psi_3^{(0)}\xi_i^{(0)})r^2 dx^i, \\
m=-1/2\Psi_3^{(0)} r^2 dv+(1-\mu^{(0)} r)\xi_i^{(0)} dx^i\\-(\bar{\lambda}^{(0)}r+1/2\bar{\Psi}_4^{(0)}r^2)\bar{\xi}_i^{(0)} dx^i,
        \end{aligned} \right.
\end{equation}
where $\mu$ and $\lambda$ are the spin coefficients in the Newman-Penrose formalism. Note that $\xi_i^{(0)}$ are only functions of $(x^1,x^2)$ satisfying $\xi_i^{(0)} \xi_{(0)}^i=0$ and $\xi^{(0)}_i \bar{\xi}^i_{(0)}=1$.

Following the idea in Ref.\cite{kaul1}, we choose an appropriate set of co-tetrad fields which are compatible with the metric (\ref{39}) as
\begin{equation}\begin{split}\label{5}
    e^0=\sqrt{\frac{1}{2}}(\alpha n+\frac{1}{\alpha} l),\,e^1=\sqrt{\frac{1}{2}}(\alpha n-\frac{1}{\alpha} l),\\
    e^2=\sqrt{\frac{1}{2}}(m+\bar{m}),\, e^3=i\sqrt{\frac{1}{2}}(m-\bar{m}),
\end{split}\end{equation}
where $\alpha(x)$ is an arbitrary function of the coordinates. Each choice of $\alpha(x)$ characterizes a local Lorentz frame in the plane $\mathcal{I}$ formed by $\{e^0,e^1\}$.
Restricted to the horizon $\Delta$, the revelent co-tetrad fields (\ref{5}) are given by
\begin{equation}\label{5a}\begin{split}
    e^0\triangleq e^1\triangleq \sqrt{1/2}\alpha n,\\e^2\triangleq \sqrt{2}Re(\xi_i^{(0)}) dx^i,\, e^3\triangleq -\sqrt{2}Im(\xi_i^{(0)}) dx^i.
\end{split}\end{equation}
Hereafter we denote equalities on $\Delta$ by the symbol $\triangleq$.
Notice that the nonvanishing solder fields $\Sigma^{IJ}$ on $\Delta$ satisfy
\begin{equation}\label{6}\begin{split}
\Sigma^{0i}\triangleq \Sigma^{1i},\forall i=2,3, \\
    \Sigma^{23}= i \bar{m}\wedge m\triangleq -2Im(\xi_1^{(0)} \bar{\xi}_2^{(0)} )dx^1\wedge dx^2.
\end{split}\end{equation}
By a straightforward calculation, we can get the following properties for the connection restricted to $\Delta$:
\begin{equation}\label{6b}
    A^{01}\triangleq \tilde{\kappa} dv+d (ln\alpha)\equiv d\beta(x),\,A^{0i}\triangleq A^{1i},\forall i=2,3,
\end{equation}
where $\beta(x)=\tilde{\kappa} v+ ln \alpha(x)$.

By Eqs.(\ref{6}) and (\ref{6b}) the horizon integral can be reduced to
\begin{equation}\label{10}
\frac{1}{\kappa} \int_{\Delta}\delta_{[1}(*\Sigma)_{IJ}\wedge \delta_{2]} A^{IJ}=\frac{2}{\kappa}\int_{\Delta} \delta_{[1}\Sigma^{23}\wedge\delta_{2]} A^{01}.
\end{equation}
In fact, $(*\Sigma)_{01}=\Sigma^{23}$ is the area element 2-form on the slicing $v=const.$ of the horizon, since the property of an isolated horizon ensures that the area of the slice is unchanged for different $v$. We can conclude that
\begin{equation}\label{6e}
    d(*\Sigma)_{01}=d\Sigma^{23}\triangleq 0.
\end{equation}
Thus $\Sigma^{23}$ is closed. So we can define a 1-form $\tilde{B}$ locally such that
 \begin{equation}\label{6d}
  \Sigma^{23}=d\tilde{B} .
\end{equation}
 Note that the topology of the horizon $\Delta$ is nontrivial with the second cohomology group $H^2(R\times S^2)\cong \mathbb{R}$.
Hence the $\tilde{B}$ field cannot be globally defined on $\Delta$. This situation is similar to the monopole in electromagnetism theory. Although there is no globally defined potential for the electromagnetic field in a topologically nontrivial spacetime, one can define the so-called Wu-Yang potential\cite{wy1} for separated  topologically trivial regions. Indeed, we have the following condition for the integral over any cross section of $\Delta$:
\begin{equation}\label{8}
    \oint_{S^2} d\tilde{B}=\oint_{S^2} \Sigma^{23}=-\oint _{S^2}2Im(\xi_1^{(0)} \bar{\xi}_2^{(0)} )dx^1\wedge dx^2=a_H.
\end{equation}
where $a_H$ represents the area of the horizon.

 Consider a SO(1,1) boost on the plane spanned by $\{e^0,e^1\}$ with group element $g=\exp(\zeta)$. Under this transformation, $A^{'01}=A^{01}-d\zeta$ and $\Sigma'_{23}=\Sigma_{23}$ are unchanged. Hence $A^{01}$ is a SO(1,1) connection, and $\Sigma^{23}$ is in its adjoint representation. We will see later that this is just what we need for a SO(1,1) BF theory.

Inserting Eqs.(\ref{6b}) and (\ref{6d}) into Eq.(\ref{10}), we get
\begin{equation}\label{11}\begin{split}
    \int_{\Delta}\delta_{[1}\Sigma^{23}\wedge\delta_{2]} A^{01}=d\int_{\Delta}\delta_{[1} \Sigma^{23}\wedge \delta_{2]}\beta\\=\oint_{H_1}\delta_{[1} \Sigma^{23}\wedge \delta_{2]}\beta-\oint_{H_2}\delta_{[1} \Sigma^{23}\wedge \delta_{2]}\beta.\end{split}
\end{equation}
Note that the bulk term in Eq.(\ref{2b}) can be rewritten as the usual form in LQG\cite{el1}, with the new variables $\mathcal{A}_{\mu}^i=\gamma A_{\mu}^{0i}-1/2\epsilon^{i}_{\ jk}A^{jk}_{\mu}$, and $\Sigma^i=\epsilon^i_{\ jk}\Sigma^{jk}$.
Then the full presymplectic structure can be defined on a spatial slice $M$ with the inner boundary $H=M\bigcap \Delta$ as
\begin{equation}\label{11a}\begin{split}
    \Omega(\delta_1,\delta_2)=\frac{1}{2\kappa\gamma}\int_M 2\delta_{[1}\Sigma^i\wedge\delta_{2]}\mathcal{A}_i+\frac{1}{\kappa}\oint_{H}2\delta_{[1} \Sigma^{23}\wedge \delta_{2]}\beta\\
\equiv \Omega_M(\delta_1,\delta_2)+\Omega_H(\delta_1,\delta_2),
\end{split}\end{equation} which is independent of the choice of the spatial surface $M$.
As we can see, the presymplectic form is split into the bulk term and the boundary term. Hence we can handle the quantization of the bulk and boundary degrees of freedom separately. In the following section, we will show that the presymplectic form on the boundary is precisely that of a topological SO(1,1) BF theory  with locally defined $B$ fields on the isolated horizon.

\section{Three-Dimensional SO(1,1) BF Theory}
In a three-dimensional spacetime $\Sigma$ without boundary, the action of the SO(1,1) BF theory can be written as\cite{bf1,rov2}
\begin{equation}\label{12}
    S[B,A]=\int_{\Sigma} B\wedge F(A)=\int_{\Sigma} dB\wedge A.
\end{equation}
Since one has $SO(1,1)\cong \mathbb{R}$, the connection field $A$ is a real-valued 1-form, and the $B$ field is also a real-valued 1-form.
From the action (\ref{12}), we can easily get the equation of motion as
\begin{equation}\label{12a}
    F=dA=0,\quad dB=0.
\end{equation}
In the Hamiltonian formalism, the restriction of the fields $A$ and $B$ to the spatial hypersurface gives the conjugate variables, which we still denote as $(A,B)$, satisfying the Gaussian constraint as well as the constraint $F=dA=0$\cite{bf1}. The latter generates gauge transformations of the form
\begin{equation}\label{12b}
    A\rightarrow A,\quad B \rightarrow B+d \lambda.
\end{equation}
From the viewpoint of covariant phase space, the symplectic flux can be obtained from the antisymmetrization of the second variation of action (\ref{12}) as
\begin{equation}\label{13}
\int_{\Sigma} 2\delta_{[1}(dB)\wedge\delta_{2]} A,
\end{equation}
from which we can get the presymplectic form on the covariant phase space as\cite{mm1}
\begin{equation}\label{13a}
    \Omega(\delta_1,\delta_2)=\oint_{\tilde{H}} 2\delta_{[2}B\wedge \delta_{1]}A,
\end{equation} where $\tilde{H}$ is an arbitrary two-dimensional spatial slice in $\Sigma$.

It should be noted that if $\tilde{H}$ is topologically nontrivial and the $B$ field is not globally defined (as was the case in the last section) the definition of the integration in the presymplectic form (\ref{13a}) is a delicate issue. However--as shown in the Appendix--in the case of a two-sphere $\tilde{H}=S^2$, the integration can be well defined as the sum of integrals over two topological trivial patches and one of their boundaries. Then the boundary presymplectic form $\Omega_H$ in Eq.(\ref{11a}) can be regarded as that of SO(1,1) BF theory by making the identification
\begin{equation}\label{14}
    B\leftrightarrow \frac{\tilde{B}}{\kappa},\,A\leftrightarrow A^{01}.
\end{equation}
Hence on the nonrotating isolated horizon, the boundary degrees of freedom of general relativity can be described effectively by a SO(1,1) BF theory.
Since the fundamental group of the manifold $\Delta$ is trivial, i.e, $\pi_1(\mathbb{R}\times S^2)$ is trivial, the quantum BF theory has trivial Hilbert space\cite{wit2,car2}.

Recall that in canonical LQG, the kinematical Hilbert space is spanned by spin network states $|\Gamma,\{j_e\},\{i_v\}>$\cite{ash0,ma1}, where $\Gamma$ denotes some graph in the spatial manifold $M$, each edge $e$ of $\Gamma$ is labeled by a half-integer $j_e$ and each vertex $v$ is labeled by an intertwiner $i_v$. In the case when $M$ has a boundary $H$, some edges of $\Gamma$ may intersect $H$ and endow it with a quantum area at each puncture\cite{enpp1}. Thus, to account for the isolated horizon degrees of freedom, we need to consider the quantum BF theory with sources.
Equations (\ref{6b}) and (\ref{6d}) imply that the equation of motion of our BF theory is
\begin{equation}\label{14a}
   F=dA=0,\quad dB=\frac{\Sigma^1}{2\kappa}.
\end{equation}
Comparing with Eq.(\ref{12a}), Eq.(\ref{14a}) shows that the bulk field $\Sigma^1$ is the source of the $B$ field rather than the $A$ field. In the Hamiltonian formalism, the constraint $F=0$ still generates the gauge transformation (\ref{12b}), and the Gaussian constraint $dB=\frac{\Sigma^1}{2\kappa}$ contains the source term but still generates the SO(1,1) gauge transformation,
\begin{equation}\label{14aa}
    A\rightarrow A+d\lambda, \quad B\rightarrow B.
\end{equation}
These two constraints form a closed algebra and hence are of first class.

Let us assume that the graph $\Gamma$ underlying a spin network state intersects $H$ by $n$ punctures denoted by $\mathcal{P}=\{p_i|i=1,\cdots,n\}$. For every puncture $p_i$ we associate a bounded neighborhood $s_i$ which contains it and does not intersect any other. We denote the boundary of $s_i$ by $\eta_i$ . Since $H$ is a homeomorphism to a two-sphere, the holonomy of flat connections is trivial. Taking account of the gauge transformations (\ref{12b}), the physical degrees of freedom of our BF theory are encoded in the flux functions
\begin{equation}\label{14b}
    f_i=\int_{s_i}dB=\oint_{\eta_i}B ,
\end{equation}
which are gauge-invariant functions of $B$.
Since we can associate a real-valued variable $f_i$ to each puncture $p_i$ , the configuration space of the BF theory with $n$ punctures is $\mathbb{R}^n$. Therefore, we can employ the well-known Lebesque measure to define the quantum Hilbert space $\mathcal{H}_H^{\mathcal{P}}$ as the space of $L^2$ functions on $\mathbb{R}^n$. Note that, as configuration operators, $\hat{f}_i$ act on any wave function by multiplications. The common eigenstates of all these $\hat{f}_i$ are the Dirac distributions $(\{a_p\},\mathcal{P}|\equiv (a_1,a_2,\cdots,a_n|$ characterized by $n$ real numbers $\{a_i,i=1,\cdots,n\}$. As unbounded self-adjoint operators, the collection $\{\hat{f}_i| i=1,\cdots,n\}$ comprises a complete set of observables in $\mathcal{H}_H^{\mathcal{P}}\equiv L^2(\mathbb{R}^n)$. There is a spectral decomposition of $\mathcal{H}_H^{\mathcal{P}}$ with respect to each $\hat{f}_i$, i.e,
\begin{equation}\label{15}
    (\{a_p\},\mathcal{P}|\hat{f}_i=(\{a_p\},\mathcal{P}|a_i.
\end{equation}

\section{Boundary Condition and State Counting}
The form of the presymplectic form (\ref{11a}) motivates us to handle the quantization of the bulk and horizon degrees of freedom separately. As in the standard LQG one first considers the bulk kinematical Hilbert space $\mathcal{H}^\mathcal{P}_M$ defined on a graph $\Gamma \subset M$ with the $n$ punctures $\mathcal{P}$ as the end points on $H$. This Hilbert space can be spanned by the spin network states $|\mathcal{P},\{j_p,m_p\};\cdots>$, where $j_p$ and $m_p$ are, respectively, the spin labels and magnetic numbers of the edge $e_p$ with end point $p\in \mathcal{P}$. Note that the integral $\Sigma^1(H)=\int_H \Sigma^1$ can be promoted as an operator $\hat{\Sigma}^1(H)$ in $\mathcal{H}_H^{\mathcal{P}}$, and $|\mathcal{P},\{j_p,m_p\};\cdots>$ are common eigenstates of $\hat{\Sigma}^1(H)$ and the horizon area operator $\hat{a}_H$ from the viewpoint of bulk LQG. Thus we have \cite{ash0,ma1}
\begin{equation}\label{27}\begin{split}
    \hat{a}_H |\mathcal{P},\{j_p,m_p\};\cdots>\\=8\pi\gamma l^2_{Pl} \sum_{p=1}^n \sqrt{j_p(j_p+1)}|\mathcal{P},\{j_p,m_p\};\cdots >,
\end{split}\end{equation}
and \cite{enpp1}
\begin{equation}\label{28}\begin{split}
    \hat{\Sigma}^1(H)|\mathcal{P},\{j_p,m_p\};\cdots>\\=16\pi\gamma l^2_{Pl} \sum_{p\in \Gamma\cap H}m_p|\mathcal{P},\{j_p,m_p\};\cdots>.
\end{split}\end{equation}

Classically, the restriction of Eq.(\ref{14a}) to the spatial slice $H=\Delta\cap M$ implies the following boundary condition to relate the boundary and bulk degrees of freedom:
\begin{equation}\label{26}
    dB \circeq \frac{\Sigma^1}{2\kappa},
\end{equation}
where $\circeq$ means equal on $H$. Note that the constraint (\ref{26}) is of first class even for the coupled system of bulk gravity and boundary BF theory.
Hence, Eq.(\ref{26}) motivates us to use the quantum version of the horizon boundary condition as
\begin{equation}\label{29}
    (Id\otimes \hat{f}_i(s_i)-\frac{\hat{\Sigma}^1(s_i)}{2\kappa}\otimes Id)(\Psi_v \otimes \Psi_b)=0,
\end{equation}
where $s_i$ is the neighborhood of an arbitrary puncture $p_i \in \mathcal{P},\,\Psi_v \in \mathcal{H}_M^{\mathcal{P}}$, and $\Psi_b \in \mathcal{H}_H^{\mathcal{P}}$.
For a given bulk spin network state $|\mathcal{P},\{j_p,m_p\};\cdots>$, the solutions of Eq.(\ref{29}) restrict the generalized eigenstates of $\hat{f}_p$ to be $(\{m_p\}, \mathcal{P}|$ with eigenvalues

\begin{equation}\label{31a}
    a_p=\gamma m_p.
\end{equation}
 This means that by applying the quantum boundary condition, the eigenvalues $a_p$ of $\hat{f}_p$ for all $p \in \mathcal{P}$ can take values only in the subset of the real numbers consisting of the integers times a constant. Thus the quantum boundary condition not only relates the bulk and the boundary theories, but also reduces the dimension of the boundary Hilbert space.

The space of kinematical states on a fixed graph $\Gamma$, satisfying the boundary condition, can be written as
\begin{equation}\label{31}
    \mathcal{H}_{\Gamma}=\bigoplus_{\{j_p,m_p\}_{p\in\Gamma\cap H}}\mathcal{H}^{\mathcal{P}}_M(\{j_p,m_p\}) \otimes \mathcal{H}^{\mathcal{P}}_H(\{m_p\}),
\end{equation}
where $\mathcal{H}^{\mathcal{P}}_H(\{m_p\})$ denotes the subspace corresponding to the spectrum $\{m_p\}$ in the spectral decomposition of the BF theory Hilbert spaces $\mathcal{H}^{\mathcal{P}}_H$ with respect to the operators $\hat{f}_p$ on the boundary.

It should be noted that the imposition of the diffeomorphism constraint implies that one only needs to consider the diffeomorphism equivalence class of quantum states. Hence, in the following state counting, we will only take account of the number of punctures on $H$, while the possible positions of punctures are irrelevant.

To calculate the entropy of the isolated horizon that we are considering, we will follow the viewpoint of LQG to trace out the degrees of freedom corresponding to the bulk, but we will also take account of the horizon degrees of freedom\cite{ash1}. Then the entropy will be
\begin{equation}\label{32a}
  S=\ln(\mathcal{N}),
\end{equation}
where $\mathcal{N}$ is the dimension of the horizon Hilbert space compatible with the given macroscopic horizon area $a_H$ and that satisfies the horizon boundary constraint (\ref{31a}).

Now how to define the area operator of the horizon is a delicate issue\cite{blv}. In the original treatment\cite{ash1}, one employed the standard area operator (\ref{27}) defined in the kinematical Hilbert space of LQG. However, for the bulk Hilbert space $\mathcal{H}^{\mathcal{P}}_M$ with a horizon boundary $H$, the flux-area operator $\hat{a}_H^{flux}$ corresponding to the classical area $\int_H |dB|$ of $H$ can also be naturally well defined as\cite{blv}
\begin{equation}\label{35}\begin{split}
    \hat{a}_H^{flux}|\mathcal{P},\{j_p,m_p\};\cdots>=a^{flux}(\{m_p\})|\mathcal{P},\{j_p,m_p\};\cdots>
\end{split}\end{equation}
where
\begin{equation}\label{36}
    a^{flux}(\{m_p\})=8\pi\gamma l^2_{Pl}\sum_{p=1}^n |m_p|.
\end{equation}
With this choice, we have the area constraint
 \begin{equation}\label{36a}
  \sum_{p \in \mathcal{P}} |m_p|=a,\quad m_p \in \mathbb{N}/2,
\end{equation}
 where $a=\frac{a_H}{8\pi \gamma l_{Pl}^2}$.
Hence, for a given horizon area $a_H$, Eq.(\ref{31a}) implies that the horizon states satisfying the boundary condition can be labeled by sequences $(m_1,\cdots,m_n)$ subject to the constraint (\ref{36a}), where $2m_i$ are integers. As in the usual treatment in LQG\cite{ash1}, we assume that for each given ordering sequence $(m_1,\cdots,m_n)$, there exists at least one state in the bulk Hilbert space of LQG, which is annihilated by the Hamiltonian constraint. Then the dimension of the horizon Hilbert space compatible with the given macroscopic horizon area can be calculated as
\begin{equation}\label{37}
    \mathcal{N}=\sum_{n=0}^{n=2a-1} C_{2a-1}^n 2^{n+1}=2\times 3^{2a-1},
\end{equation}where $C_i^j$ are the binomial coefficients.
So the entropy is given by
\begin{equation}\label{38}
    S=\ln\mathcal{N}=2a\ln3+\ln \frac{2}{3}=\frac{\ln3}{\pi\gamma}\frac{a_H}{4l^2_{Pl}}+\ln \frac{2}{3}.
\end{equation}
Thus we have the entropy for an arbitrary nonrotating isolated horizon, which is proportional to its area at leading order. If we fix the value of the Barbero-Immirzi parameter as $\gamma=\ln 3/\pi$, the Bekenstein-Hawking area law is obtained.

It should be noted that the choice of the flux-area operator (\ref{35}) is necessary in order to get a consistent result for state counting. Had we chosen the area operator (\ref{27}) in full LQG to represent the horizon area \cite{ash1}, we would have the area constraint
\begin{equation}\label{32}
    8\pi \gamma l_{Pl}^2\sum_{p=1}^n \sqrt{j_p(j_p+1)} = a_H.
\end{equation}
On the other hand, there is a global constraint which follows from the quantum versions of Eqs.(\ref{8}) and (\ref{31a}),
\begin{equation}\label{33}
    \sum_{p\in \mathcal{P}} |a_p|=\gamma \sum_{p\in \mathcal{P}} |m_p|=a_H/\kappa.
\end{equation}
Since $m_i\in \{-j_i,\cdots,j_i\}$, there is no common solution for the both of the constraints.

\section{Discussion}
In the previous sections, the nonrotating isolated horizon in four-dimensional spacetime has been studied, and its entropy has been calculated in the framework of LQG. By the gauge choice of Eq.(\ref{5}), the degrees of freedom of the horizon can be encoded in a SO(1,1) BF theory. From the view point of LQG, the spin networks of the bulk quantum geometry puncture the horizon, endowing it with quantum area. This picture not only transforms the horizon boundary condition (\ref{26}) into the quantum condition (\ref{29}), but also indicates the area constraint (\ref{36a}). Thus, for a given macroscopic horizon area, the microscopic degrees of freedom of the horizon can be calculated as in Eq.(\ref{37}), which accounts for its entropy and suggests a value for the Barbero-Immirzi parameter.

It should be noted that these microscopic degrees of freedom on the horizon boundary are all of quantum nature. Although in classical theory the boundary has no independent degrees of freedom, because of the distributional nature of the connection and flux operators in quantum theory, they can fluctuate ay the boundary but have to obey the boundary condition. Recall that in the Chern-Simons theory description of the horizon \cite{el1}, the boundary degrees of freedom are encoded in the Chern-Simons connection. However, in our BF theory description, as we can see from Eq.(\ref{14a}), the connection becomes pure gauge, while the nontrivial degrees of freedom of the boundary are all encoded in the $B$ field.

Note also that our starting point is the Palatini action (\ref{1}). If the Immirzi parameter term was added in the action, the new connection variable would not satisfy Eq.(\ref{6b}). Then it would be difficult to derive a BF symplectic form on the horizon boundary. In our treatment, the Immirzi parameter is introduced through the canonical transformation below Eq.(\ref{11}). Compared with the Chers-Simons theory approach, our BF theory approach indicates a different value of the Barbero-Immirzi parameter. However our value for the parameter coincides with its value obtained in a particular case in Ref.\cite{blv} by employing the same flux-area operator as ours but in the approach of Chern-Simons theory. Whether this coincidence implies any relation between the two approaches deserves further investigation. The quantum correction to the Bekenstein-Hawking area law in our approach is a constant, $\ln(2/3)$, while the Chern-Simons theory approach usually gives a logarithmic correction at first order. This delicate issue of the quantum correction to the classical area law of the isolated horizon was also discussed in Ref.\cite{m1}. Irrespective of these differences, by taking account of the fact that LQG can be extended to arbitrary spacetime dimensions\cite{btt2,btt3,btt4,btt5,btt1} the virtue of our BF theory approach is that it admits an extension to an arbitrary-dimensional horizon\cite{wm1}, while the Chern-Simons theory can only live on odd-dimensional manifolds.

\begin{acknowledgments}
The authors would like to thank Alejandro Perez for helpful discussion. This work is supported by the NSFC (Grant No. 11235003) and the Research Fund for the Doctoral Program of Higher Education of China.
\end{acknowledgments}

\section{Appendix}
We now show that the boundary presymplectic form in Eq.(\ref{11a}) is indeed that of a BF theory with locally defined $B$ fields on the horizon.

Note that the $B\equiv \tilde{B}$ field is defined locally through Eq.(\ref{6d}). We can cover the two-sphere $S^2$ with two topologically trivial patches $S_+$ and $S_-$ with boundaries $c_1$ and $c_2$, respectively. The intersection region is denoted by $S_0=S_+\cap S_-$ with boundary $c_1+c_2$. The potentials in the regions $S_+$ and $S_-$ can be separately well defined as $B_+$ and $B_-$, satisfying $dB_+=dB_-=\Sigma^{23}$. In the region $S_0$, we have both $B_+$ and $B_-$ such that $B_+-B_-=g$, where $g$ is a closed 1-form. So we have
\begin{widetext}\begin{equation}\label{51}\begin{split}
    \oint_{S^2}\delta_1 \Sigma^{23} \wedge \delta_2 \beta=\int_{S_+}\delta_1 (dB_+) \wedge \delta_2 \beta+\int_{S_-}\delta_1 (dB_-) \wedge \delta_2 \beta-\int_{S_0}\delta_1 (dB_+) \wedge \delta_2 \beta\\=\int_{S_+}\delta_1 B_+ \wedge \delta_2 A^{01}+\oint_{c_1}\delta_1 B_+ \wedge \delta_2 \beta+\int_{S_-}\delta_1 B_- \wedge \delta_2 A^{01}\\+\oint_{c_2}\delta_1 B_- \wedge \delta_2 \beta-\int_{S_0}\delta_1 B_+ \wedge \delta_2 A^{01}-\oint_{c_1+c_2}\delta_1 B_+ \wedge \delta_2 \beta\\=\int_{S^2-S_-}\delta_1 B_+ \wedge \delta_2 A+\int_{S_-}\delta_1 B_- \wedge \delta_2 A-\oint_{\partial(S_-)}\delta_1 g \wedge \delta_2 \beta,
\end{split}\end{equation}\end{widetext}
where we used the Leibniz rule, Stokes' theorem, and the definition $d\beta=A^{01}$.
Then we need to show that Eq.(\ref{51}) can be understood as the presymplectic form for a BF theory with locally defined $B$ fields such that $dB=\Sigma^{23}$ and $A(x)=d\beta(x)$.
Since the $B$ fields cannot be globally defined on $S^2$, the integration of the presymplectic form (\ref{13a}) has to be defined carefully. An innocent definition could be
\begin{equation}\label{52}\begin{split}
    \oint_{S^2}\delta_1 B \wedge \delta_2 A:=\int_{S_+}\delta_1 B_+ \wedge \delta_2 A+\int_{S_-}\delta_1 B_- \wedge \delta_2 A\\-\int_{S^0}\delta_1 B_+ \wedge \delta_2 A.
\end{split}\end{equation}
However, since both $B_+$ and $B_-$ are on the same footing in the region $S_0$, one may also employ $B_-$ instead of $B_+$ in the last integration of Eq.(\ref{52}). Obviously the two formulas are not equivalent to each other. Actually, we have
\begin{equation}\label{53}
    \int_{S_0}\delta_1 (B_+-B_-) \wedge \delta_2 A=\int_{S_0}\delta_1 g \wedge \delta_2 d\beta=-\oint_{c_1+c_2}\delta_1 g \wedge \delta_2 \beta,
\end{equation}
and hence
\begin{equation}\label{54}\begin{split}
    \int_{S_0}\delta_1 B_+ \wedge \delta_2 A+\oint_{c_2}\delta_1 (B_+ - B_-) \wedge \delta_2 \beta\\=\int_{S_0}\delta_1 B_- \wedge \delta_2 A+\int_{c_1}\delta_1 (B_- - B_+) \wedge \delta_2 \beta.
\end{split}\end{equation}
Therefore, the reasonable definition for the presymplectic form (\ref{13a}) with locally defined $B$ fields, which is independent of the choice between $B_+$ and $B_-$, should be
\begin{equation}\label{55}\begin{split}
  \oint_{S^2}\delta_1 B \wedge \delta_2 A:=\int_{S_+}\delta_1 B_+ \wedge \delta_2 A+\int_{S_-}\delta_1 B_- \wedge \delta_2 A\\-\int_{S^0}\delta_1 B_+ \wedge \delta_2 A-\oint_{c_2}\delta_1 (B_+ - B_-) \wedge \delta_2 \beta\\=\int_{S^2-S_-}\delta_1 B_+ \wedge \delta_2 A+\int_{S_-}\delta_1 B_- \wedge \delta_2 A-\oint_{\partial(S_-)}\delta_1 g \wedge \delta_2 \beta,
\end{split}\end{equation}
which coincides with Eq.(\ref{51}).

 \bibliography{btz2}
\end{document}